 \documentstyle[12pt]{article}
\begin{document}
\framebox[3.75in][l]
{To appear in {\it Philosophical Magazine}, 1996.}
\vspace*{0.25in}
 \begin{center} \large Low-Energy Excitations in an
  Incipient Antiferromagnet  \\
  \end{center}
\vspace{0.2in}
 \begin{center}
 {\em D.W. Hess} \\
 Complex Systems Theory Branch \\
 Naval Research Laboratory \\
 Washington, D.C. \ \ 20375-5345
 \end{center}
\vspace{0.05in}
 \begin{center}
  {\em J.J. Deisz} and {\em J.W. Serene} \\
   Department of Physics \\
   Georgetown University \\
   Washington, D.C. \ \ 20057-0995
 \end{center}

\vspace{0.15in}

\begin{center}
 \parbox{4.5in}{{\bf Abstract:}
 We present fully self-consistent calculations in the
 fluctuation exchange approximation for
 the half-filled Hubbard model in 2D.
 A non-fermi liquid state evolves with decreasing
 temperature in this self-consistent
 model of coupled spin fluctuations
 and quasiparticles.  The mean field phase transition to
 long-range antiferromagnetic order is suppressed and
 we find no evidence of a phase transition to
 long-range magnetic order. We show that
 the real part of the self-energy at zero energy
 shows a positive slope and the imaginary part of
 the self-energy shows a local minimum. The scale of
 this structure is set by the zero temperature
 gap in mean field theory.  The growth of  spin
 fluctuations is reflected in the evolution of
 sharply peaked structure in the spin fluctuation
 propagator around zero energy and ${\bf Q} = (\pi, \pi)$.
 We present calculations for the Hubbard model in 1D in
 a two-particle self-consistent parquet approximation.
 A second moment sum rule suggests
 that vertex corrections to the self-energy are responsible
 for the greater accuracy of the parquet approximation as
 compared to other self-consistent perturbation theories.
 }
\end{center}

 \begin{center}
{\large {\it \S 1 Introduction and Physical Picture}}
 \end{center}

  The role of spin fluctuations in the physics of a many-body
  Fermi system seems a particularly appropriate topic for a
  festschrift in honor of David Pines.  In liquid $^3$He,
  ferromagnetic spin fluctuations play a central role in the
  normal and superfluid states.  The exchange of spin fluctuations
  contributes a non-analytic correction to the
  quasiparticle energy that is reflected in finite-temperature
  corrections to the transport properties and
  specific heat predicted by Landau Fermi liquid theory
  (Baym and Pethick, 1991).  Similar corrections were also found
  for metallic systems near a ferromagnetic instability
  (Berk and Schrieffer 1966, Doniach and Englesberg 1966).
  In both heavy fermion materials and the high-temperature
  superconductors, the close proximity of antiferromagnetic
  instabilities and the observation of unusual and possibly
  unconventional superconductivity has aroused interest in
  the role of antiferromagnetic spin fluctuations, both as an
  explanation of the unusual normal state properties and as a
  pairing mechanism.

   The normal-state properties of the high-temperature
   superconductors have inspired speculations that interplay among
   correlations, low dimensionality and the tendency to magnetic order
   could give rise to low-energy excitations that are not adequately
   described by the Landau theory of a Fermi liquid.   Varying degrees of
   departure from Landau theory have been proposed (for recent reviews,
   see Kampf (1994) and Dagatto (1994)), ranging from the
   `soft singularities' of phenomenological marginal Fermi liquids
   and nearly antiferromagnetic Fermi liquids to more dramatic
   conjectures of separation of charge and spin excitations such as
   Luttinger liquids (Anderson, 1990).
   All the proposed departures from a Fermi liquid
   description of low energy excitations ultimately rest on subtle
   features of interactions between renormalized quasiparticles.
   Thus a self-consistent approach seems necessary for a
   complete and satisfactory theory.
   The observation of an antiferromagnetically ordered state in the
   parent compounds of the high temperature superconductors suggests
   the possibility that the renormalization of quasiparticle
 excitations by the exchange of spin fluctuations may account
  for the unusual properties of these materials.
   Here we describe results for a self-consistent model aimed at
   elucidating the physics of
   coupled spin fluctuations and quasiparticle excitations.

   The 2D copper-oxide layer structure of the cuprate high temperature
   superconductors has inspired great interest in the single-band Hubbard
   Hamiltonian in 2D as a model for the high temperature superconductors
   (Anderson, 1987).
   Numerical evidence from quantum Monte Carlo calculations
  strongly indicates that the ground state of the half-filled Hubbard
  model in 2D possess long-range antiferromagnetic
  order (Hirsch and Tang, 1989; White, Scalapino, Sugar, Loh, Gubernatis,
  and Scalettar, 1989);
  at finite temperature, the possibility of long-range magnetic order is
  precluded by the Mermin-Wagner theorem.  In this paper we consider
  the 2D Hubbard model at half-filling, where spin fluctuations are
  expected to play a pivotal role as the system of
  coupled quasiparticles and (incoherent) spin fluctuations
  self-consistently evolves with decreasing
  temperature toward the ordered state at zero temperature.
  For small $U$, mean field theory predicts a phase transition
  to an antiferromagnetically ordered state at a non-zero temperature.

  In a previous work (Deisz, Hess and Serene, 1995) we showed that the
  self-consistent inclusion of fluctuations at the level of the
  fluctuation exchange
  approximation (FEA) strongly suppresses the mean field phase transition;
  we have not observed a magnetic phase transition down to
  the lowest temperatures for which we have calculations. At sufficiently low
  temperature, the coupling of growing spin fluctuations to quasiparticle
  excitations leads to the evolution of a non-Fermi liquid state, clearly
  signaled by anomalous features in the self-energy
  for $\varepsilon \sim 0$: a positive slope in the real part and
  a local minimum in the imaginary part.  Spectral weight is rapidly
  pushed away from the Fermi surface with decreasing temperature
  leading to the formation of a weak pseudogap in the single-particle
  density of states.  We presented a simple
  analytical model which showed that these features, along with
  developing shadow structure in single-particle spectral
  functions, are a consequence of evolving spin fluctuations
  and their self-consistent effect
  on the structure of evolving quasiparticles.

  This paper extends our previous results providing
  further details of the
  evolution of the non-Fermi liquid state with decreasing temperature.
  We observe that the
  structure that develops in the self-energy at low energy
  occurs on the
  scale of the zero temperature gap in mean field theory.  The formation
  of the non-Fermi liquid state is accompanied by the evolution of
  a sharply peaked structure in the $T$-matrix;
  the temperature and energy dependence of this structure suggests
  the existence of a strongly temperature dependent characteristic
  energy scale not unlike that proposed in the phenomenologically
  motivated NAFL (Pines, 1990;  Millis, Monein and Pines, 1990).
  To make this discussion self contained, we briefly summarize the FEA
  in the next section.  In \S 3 we present a short summary of our numerical
  methods; it will be convenient there to summarize some formally exact
  results that suggest that FEA is a better approximation than including
  only the sum of the particle-hole bubble diagrams (the shielded potential
  approximation).
  In \S 4 we describe the non-Fermi liquid solution
  that evolves at low temperature.  Finally in \S 5, we present
  results for the 1D Hubbard model at half filling
  in a fully self-consistent parquet approximation.
  This approximation
  includes interactions between fluctuations in the calculation of the
  self-energy and is self-consistent at the level of the two-particle
  propagator.

 \begin{center}
  {\large {\it \S 2 The Fluctuation Exchange Approximation
   for the Hubbard Model}}
 \end{center}

  In the propagator renormalized perturbation theory of
   Luttinger and Ward (Luttinger and Ward, 1960)
   the grand thermodynamic potential $\Omega$
   is constructed as an
   independent functional of the fully renormalized single particle
   propagator $G$ and the self-energy $\Sigma$, and is stationary
   with respect to independent variations of $G$ and $\Sigma$.
   Dyson's equation follows from
   \begin{equation}
   {{\delta \Omega [G, \Sigma] } \over {\delta \Sigma}} = 0,
   \label{one}
   \end{equation}
   and the representation of the self-energy as a sum of `skeleton'
   graphs evaluated with
   the fully renormalized propagator and bare two-particle vertex
   is obtained from
   \begin{equation}
   {{\delta \Omega [G, \Sigma] } \over {\delta G}} = 0.
   \label{two}
   \end{equation}
   The self-consistent solution of Eq. (\ref{one}) and Eq. (\ref{two})
   yields the exact Green's function.
   Explicit summation of the infinite set of `skeleton' graphs
   for the self-energy required for the exact solution is
   generally not possible and approximations are generated by
   selecting a class or classes of vacuum graphs
   for the grand potential.
   The FEA is one such approximation; it includes the exchange
   of density fluctuations, spin-density fluctuations and
   singlet-pair fluctuations.

  It is convenient to discuss the FEA by alternating
  between position and (imaginary) time, and momentum and
  (imaginary) frequency representations.
  The FEA for the electron self-energy of the paramagnetic state is given by
\begin{equation}
\Sigma({\bf r},\tau) = U^2 \left[ \chi_{ph}({\bf r},\tau) + T_{\rho\rho}
({\bf r},\tau) + T_{sf}({\bf r},\tau) \right] \,G({\bf r},\tau) +
U^2 T_{pp}({\bf r},\tau)G(-{\bf r},-\tau)
\label{fea}
\end{equation}
where
$\chi_{ph} = - G({\bf r},\tau)\, G(-{\bf r},-\tau)$ and
$\chi_{pp} = G({\bf r},\tau)\, G({\bf r},\tau)$ are
the particle-hole and particle-particle susceptibility bubbles.
The density, spin, and particle fluctuation $T$-matrices are
most simply written in
momentum-frequency space,
\begin{eqnarray}
T_{sf}({\bf q},\omega_m) & = &  \
 \  {{3}\over{2}} \ \ {{U \chi_{ph}({\bf q},\omega_m)^2 }\over
                             {1 - U \chi_{ph}({\bf q},\omega_m)}} \\
T_{\rho\rho}({\bf q},\omega_m) & = &
- {{1}\over{2}} \  {{U \chi_{ph}({\bf q},\omega_m)^2 }\over
                             {1 + U \chi_{ph}({\bf q},\omega_m)}} \\
T_{pp}({\bf q},\omega_m) & = & \ \ \ \ \ \
  {{U \chi_{pp}({\bf q},\omega_m)^2 }\over
                             {1 + U \chi_{pp}({\bf q},\omega_m)}}.
\label{t_matrix}
\end{eqnarray}
The Green's function is obtained from Dyson's equation
\begin{equation}
G({\bf k},\varepsilon_n)^{-1} =G_o({\bf k},\varepsilon_n)^{-1} -
\Sigma({\bf k},\varepsilon_n)
\label{dyson}
\end{equation}
where $G_o({\bf k},\varepsilon_n)$ is the non-interacting Green's
function.  The FEA self-energy (and therefore $G$) are found from
the self-consistent solution of Eqs. (\ref{fea} -- \ref{dyson}).

\begin{center}
  {\large {\it \S 3 Numerical Methods and Analytic Considerations}}
\end{center}

   We briefly describe our methods here; a more detailed account
  can be found elsewhere (Deisz, Hess and Serene, 1994).  The
  self-consistent self-energy is obtained by the iterative
  solution of Eqs.\ (\ref{fea}--\ref{dyson}).  Efficient fast Fourier
  transforms are used to switch between $({\bf r}, \tau)$-space
  and $({\bf k}, \varepsilon_n)$-space.  Since we use periodic
  boundary conditions on a periodic lattice,
  the use of discrete Fourier transform (DFT) methods introduces
  no fundamental difficulties in the transformation between ${\bf k}$
  and ${\bf r}$ variables. On the other hand, a discrete sampling of
  the $\tau$ variable generally introduces an artificial periodicity
  (with a period of the cutoff frequency)
  in the space of (discrete) Matsubara frequency.  Nonetheless, all
  currently used high frequency cutoff schemes can be simulated using DFT's,
  with the possible cost of doubling the number of frequencies used
  (Serene and Hess, 1992).  Building on the observation that
  comparatively little information
  is contained in the high frequency behavior of $G$, the $T$-matrices, the
  susceptibility bubbles, and $\Sigma$, we have devised a substantially
  more accurate method that preserves the formal internal self-consistency
  of the theory, improves the stability of the algorithm, and
  enables thermodynamic quantities to be calculated reliably.
  We decompose all quantities into the sum of a numerical term and
  an analytic term. The analytic terms are asymptotically exact at high
  frequency; they allow highly accurate calculations of contributions to
  low-frequency properties from convolution sums over high-frequency
  tails.  These contributions, which propagate through
  any iterative solution of the nonlinear equations of the FEA, are lost in
  other implementations that introduce a high frequency cutoff.
  We find that accurate treatment of these contributions is absolutely
  essential for meaningful results from the FEA in the presence of large
  fluctuations.  We perform calculations on $128 \times 128$ lattices
  of {\bf k}-points with typically 512 Matsubara frequencies.
  A fine mesh is required to represent accurately important fine structure
  that evolves, especially in the spin-fluctuation propagator,  with decreasing
  temperature.  The resulting algorithm has yielded converged
  solutions at half-filling for a large range of $U$ with continuous
  control of the temperature $T$ down to $T \sim 0.008 t$ the lowest
  temperatures we have explored.  Numerical analytic continuation
  of quantities calculated on the imaginary frequency axis to the
  real frequency axis is accomplished
  using Pad\'e approximants (Vidberg and Serene, 1977).

\begin{figure}[htb]
\vspace{4.25in}
\includegraphics{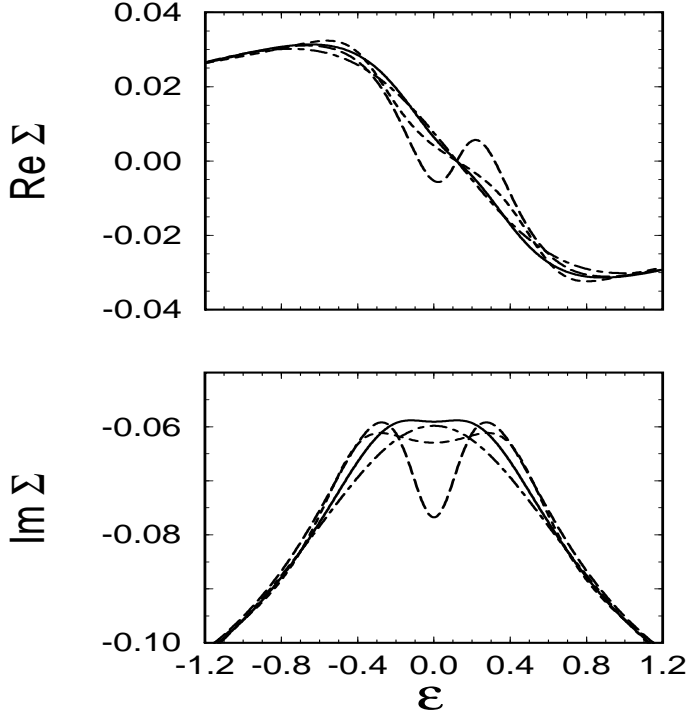}
\center{
\parbox{5.5in}{
{\small
\caption{
    The evolution of a non-Fermi liquid
    self-energy with decreasing temperature.
    Shown are $\Sigma({\bf k}_F
     = (\pi/8, 7\pi/8), \varepsilon)$
    for $T = 0.06$ (long dash),
    $0.07$ (dashed), $0.09$ (solid) and
    $0.12$ (dot-dash).  Note that above
    $\varepsilon \sim 1$,
    Im $\Sigma ({\bf k}, \varepsilon)$ displays
    the linear behavior consistent with our
    earlier observations (Serene and Hess, 1992).
}
\label{fig1}
}
}}

\end{figure}
\begin{figure}[htb]
\vspace{4.25in}
\includegraphics{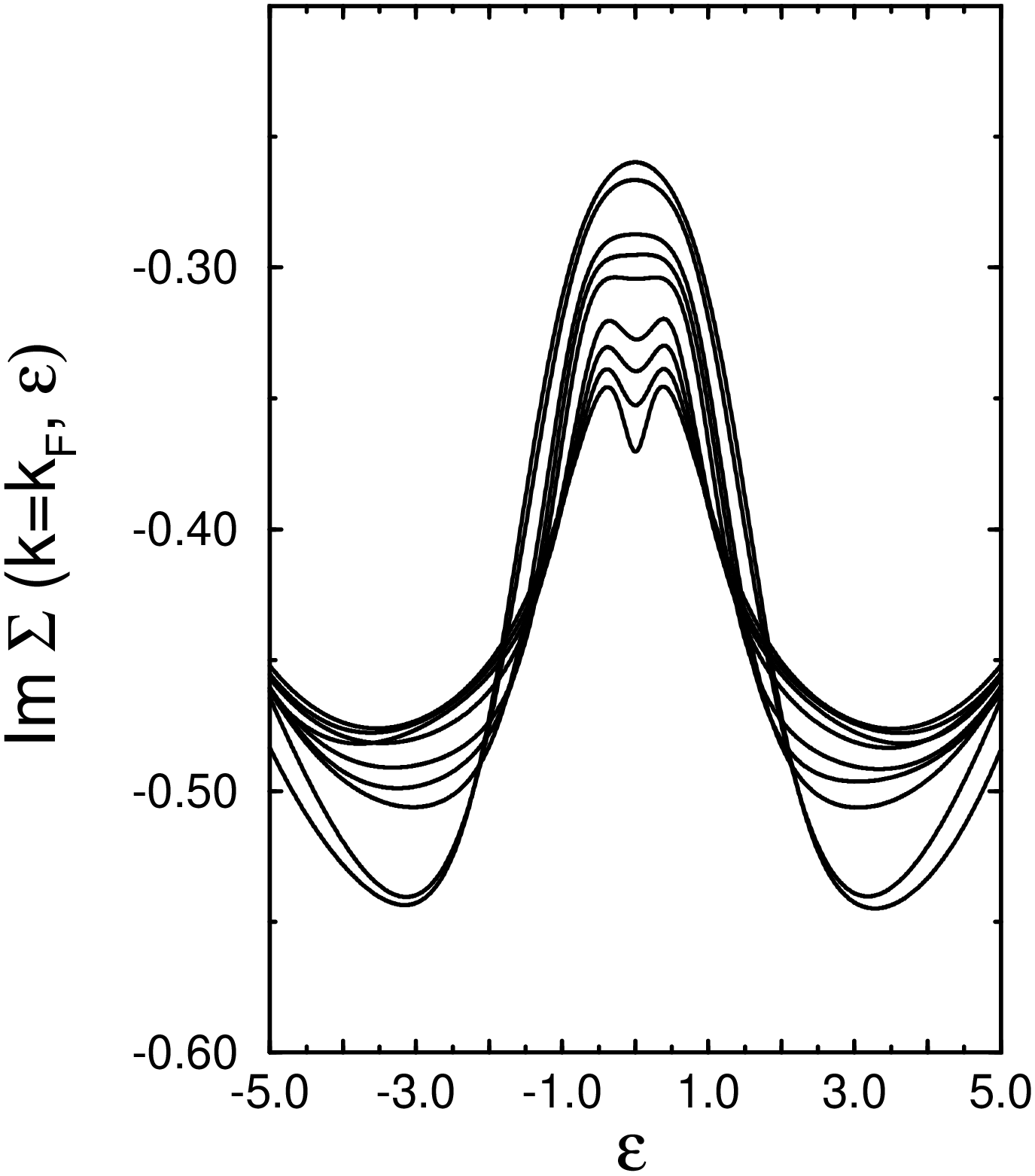}
\center{
\parbox{5.5in}{
{\small
\caption{
  Im $\Sigma({\bf k}_F, \varepsilon)$
    for $U= 2.7$ and $T = 0.10$
    showing the smooth transformation from
    clear non-fermi liquid behavior at
    ${\bf k} = (\pi, 0)$
    on the Fermi surface (lowest curve at
    $\varepsilon = 0$) to a Fermi-liquid like
    behavior at the point ${\bf k} = (\pi/2, \pi/2)$
    on the Fermi surface along the $(1,1)$
    direction (highest curve at $\varepsilon = 0$).
}
\label{fig2}
}
}}
\end{figure}

  The slowly decaying high frequency tails of the
  single-particle Green's function
  are a consequence of the discontinuity of $G$ and its derivatives
  at $\tau = 0$.  For example, the discontinuity
  $G(0^+) - G( 0^-) = -1$ ( which follows from the canonical
  anticommutation relations for Fermions) results in the
  $1/i \varepsilon_n$ behavior of $G$ at high frequency. More generally,
  discontinuities in
  $G(\tau)$ and its derivatives can be related to weighted integrals
  of the spectral function,
 \begin{equation}
   {{\partial^n G({\bf k}, \tau)} \over {\partial \tau^n}}|_{\tau=0^+}
 - {{\partial^n G({\bf k}, \tau)} \over {\partial \tau^n}}|_{\tau=0^-}
 = (-1)^{(n+1)} \int_{-\infty}^{\infty}
     \varepsilon^n A({\bf k}, \varepsilon) d \varepsilon.
 \end{equation}
  The left hand side of this equation can be evaluated directly using
  the equations of motion for $G$, yielding exact sum rules for the
  single particle spectral function.  From Dyson's equation for the
  exact $G$ and the `skeleton' diagram expansion for $\Sigma$
   one can show (Deisz, Serene and Hess, 1995)
  that for the Hubbard Hamiltonian:
  (1) the contribution from the self-energy to the $n=1$ sum rule
      comes entirely from the Hartree-Fock diagram
      with the exact $G$, and
  (2) the contribution from the self-energy to the $n=2$ sum rule
      comes entirely from the second-order `skeleton' diagram
      evaluated with the exact $G$.
  When evaluated with the FEA Green's function, the first moment sum
  rule is satisfied but the second generally is not.  It is important
  to note, however, that the violation of the sum rule is significantly
  greater for approximations that keep only particle-hole bubble
  contributions.  From this point of view,
  the FEA is a better approximation than taking particle-hole
  bubble or particle-particle ladder `skeleton' graphs alone.

\begin{center}
 {\large {\it \S 4 Neither Fermi-Liquid Nor Antiferromagnet}}
\end{center}

  We have calculated self-consistent self-energies in the fluctuation
 exchange approximation for the 2D Hubbard model at half filling.
  In Fig. 1, we show the self-energy at low energy for a momentum
  on the Fermi surface near the $X$ point and for various temperatures.
  For this modest $U = 1.57$
 (all energies are understood to be measured with respect to the
  hopping matrix element $t$) we observe
   that at high temperature  ($T \stackrel{>}{\sim} 0.12$,
   approximately the mean field transition temperature)
  the energy dependence of $\Sigma$ at low energy
  is roughly similar to that of Fermi liquid theory: the slope of
   Re $\Sigma$ corresponds to a
  quasiparticle pole weight near unity, and Im $\Sigma$ shows
  a roughly parabolic energy dependence.  It is important to
  note, however, that Im $\Sigma({\bf k}_F, \varepsilon=0)$ is
  significantly larger in magnitude compared with that calculated for an
  approximation containing only the second order diagram
  (Deisz {\em et al.}, 1995).  This
  reflects enhanced quasiparticle-quasiparticle scattering by AFM
  spin fluctuations.
  As the temperature is lowered, $\Sigma$ develops structure
  at $\varepsilon =0$ that is inconsistent with a Fermi liquid;
  the slope of Re $\Sigma$
  decreases rapidly and eventually changes sign,
  and Im $\Sigma$ develops a local minimum.
  These anomalies occur at roughly
  half the mean field theory transition temperature
  to the antiferromagnetically ordered state
  in mean field theory.
  Moreover, the energy scale suggested by the local maxima in
  Im $\Sigma$ at $\varepsilon \sim 0.28$ is very nearly the
  size of the zero temperature gap $\Delta$ in mean field theory
  obtained by the ${\bf k}$-sum over the magnetic Brillion zone,
  \begin{equation}
   U^{-1} = {{1} \over {N}}
      \sum_k {{1} \over {\sqrt{\epsilon_k^2 + \Delta^2}}}.
  \end{equation}
  We have found this near agreement to hold for several
  values of $U$ less than half of the bandwidth.  In a separate work,
  (Deisz,{\em et. al.}, 1995)
  we have shown from calculated single-particle spectral
  functions  that the evolution of the anomaly is accompanied by
  a dramatic suppression of spectral weight at the Fermi surface.
  The loss in spectral weight at the Fermi surface
  leads to a formation of a weak pseudogap in the density
  of states at $\varepsilon=0$. As shown in Fig. 2, the evolution of
  the anomaly is anisotropic; it starts at the $X$ point and spreads
  across the Fermi surface with decreasing temperature.

\begin{figure}[htb]
\vspace{4.325in}
\includegraphics{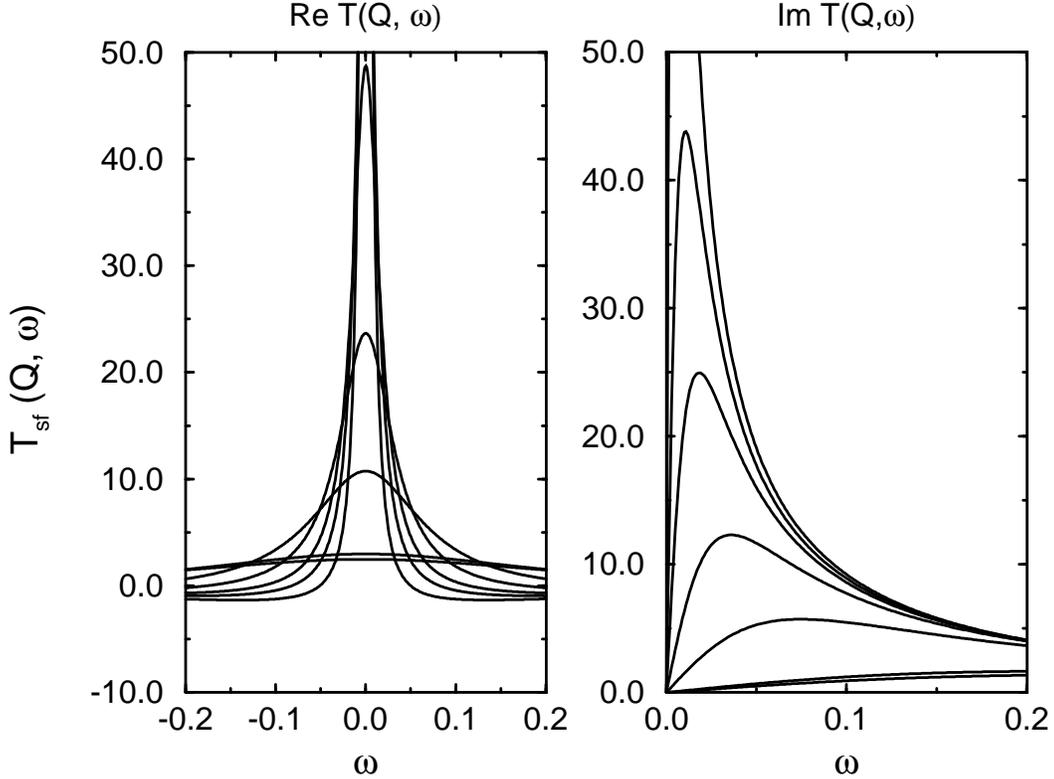}
\center{
\parbox{5.5in}{
{\small
\caption{
   The monotonic evolution with decreasing
    temperature of sharply peaked structure
    in the spin-fluctuation $T$-matrix $T_{sf}$
    on the real energy axis
    at ${\bf Q} = (\pi, \pi)$.  Shown are
    Re $T_{sf}({\bf Q}, \omega)$ and
    Im $T_{sf}({\bf Q}, \omega)$ for $T = 0.20, \, 0.18, \,
    0.10, \, 0.08, \, 0.07, \, 0.065, \, 0.06$
}
\label{fig3}
}
}}
\end{figure}

     The appearance of the non-Fermi liquid state is presaged by
  the evolution and rapid growth with decreasing temperature of structure
  in the spin-fluctuation $T$-matrix that is sharply peaked in momentum
  at $Q=(\pi, \pi)$ and in Matsubara frequency about $\omega_m = 0$.
  The analytic continuation of $T_{sf}(Q, \omega_m)$ to the real frequency
  axis is shown in Fig. 3.  The rapid increase in
  $T_{sf}(Q, \omega = 0)$ is shown in detail together with the
  temperature dependence of the width of the peak in frequency
  $\Delta_\omega T_{sf}$ and momentum $\Delta_{\bf q} T_{sf}$ (for high
  temperature, the latter quantity is anisotropic and the width
  is for the $\hat{k}_x$ direction).  Also plotted in the lower panel
  of Fig. 4 is $1/T_{sf}(Q, \omega = 0)$.  Below $T \sim 0.07$,
  $\Delta_\omega T_{sf}$ and $1/T_{sf}(Q, \omega = 0)$ track each
  other closely suggesting the existance of a characteristic temperature
  $T^*$ below which the behavior of $T_{sf}(Q, \omega)$ can be
  characterized by
  a simple diffusive pole at a (temperature dependent) characteristic
  energy $\omega^*$.  It is interesting to note that $\omega^*$
  decreases rapidly with decreasing temperature below $T^*$ and
  the admittedly small amount of data available below $T^*$ can be
  approximated well by the functional form $T_{sf}(Q, 0) = B \exp(A/T)$.
  We note that the structure of $T_{sf}({\bf q} \sim {\bf Q}, \omega)$
  below $T^*$ is similar to that of the spin-spin correlation function
  proposed by Millis, Monein and Pines, 1989.  In contrast,
  $T_{sf}({\bf q}, \omega)$ presented here is the result
  of a fully self-consistent calculation and $T_{sf}({\bf q}, \omega)$
  cannot be observed directly in an experiment; a fully self-consistent
  calculation of the dynamical spin-spin response function
  is necessary to determine how $\omega^*$ is expressed
  in measurable magnetic properties.
\begin{figure}[htb]
\vspace{5.2500in}
\includegraphics{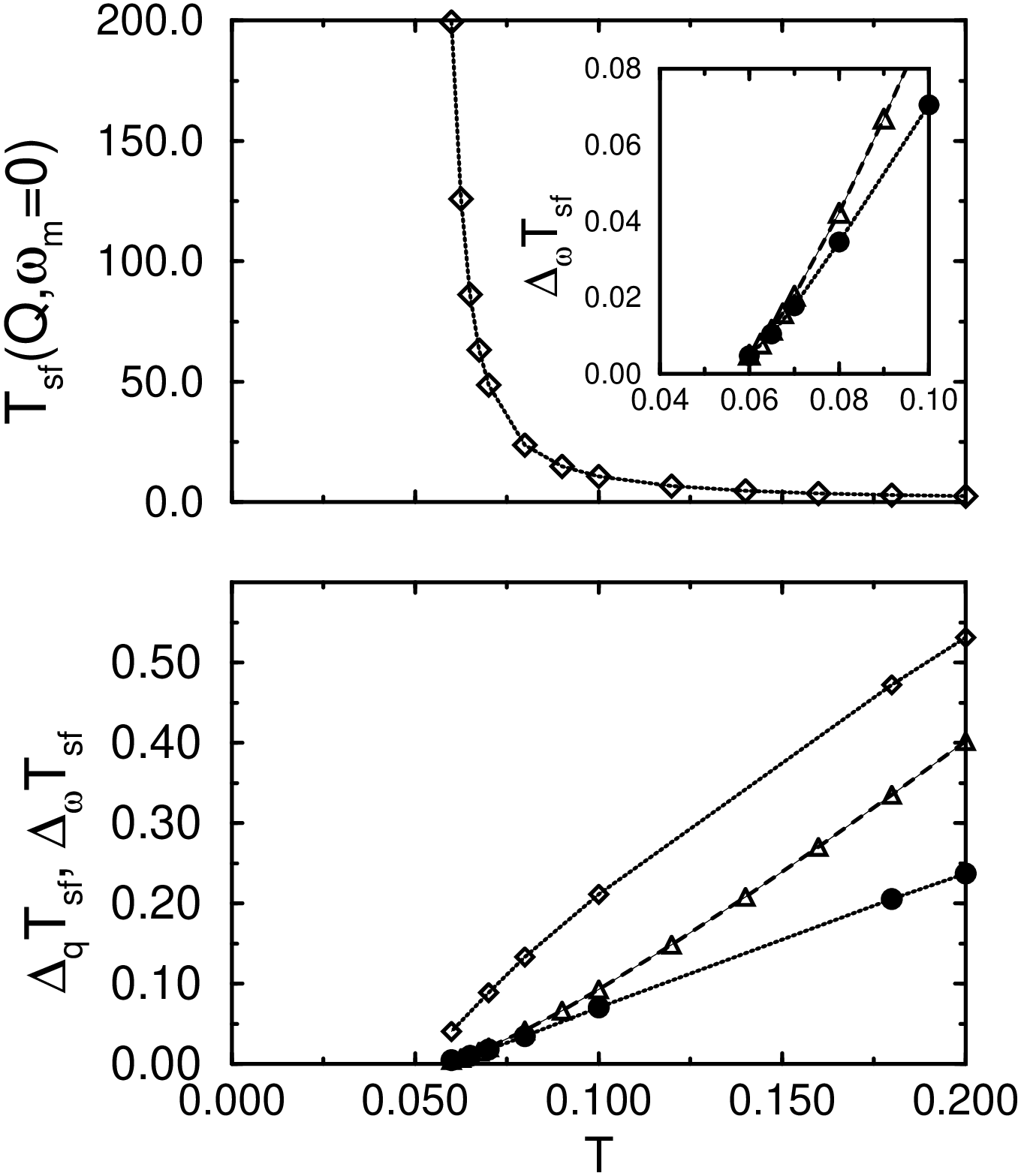}
\center{
\parbox{4.7in}{
{\small
\caption{
    The evolution of the sharply peaked
    structure of the spin fluctuation $T$-matrix
    $T_{sf}$ with decreasing temperature:
    (top panel) The maximum of $T_{sf}$
    at ${\bf Q}$ and $\omega = 0$
    as a function of T; (bottom panel) the
    temperature dependence of the
    HWHM of the sharp peak
    in $T_{sf}$ in momentum $\Delta_{\bf q} T_{sf}$
    ($\diamond$), the HWHM of $T_{sf}({\bf Q}, \omega)$
    $\Delta_{\varepsilon} T_{sf}$
    ($\bullet$), and $1/T_{sf}({\bf Q}, 0)$ ($\triangle$).
    The inset shows the merging of $\Delta_{\varepsilon} T_{sf}$
    with $1/T_{sf}({\bf Q}, 0)$
    below the temperature $T^*$ (see text).
}
\label{fig4}
}
}}
\end{figure}

\begin{figure}[htb]
\vspace{4.000in}
\includegraphics{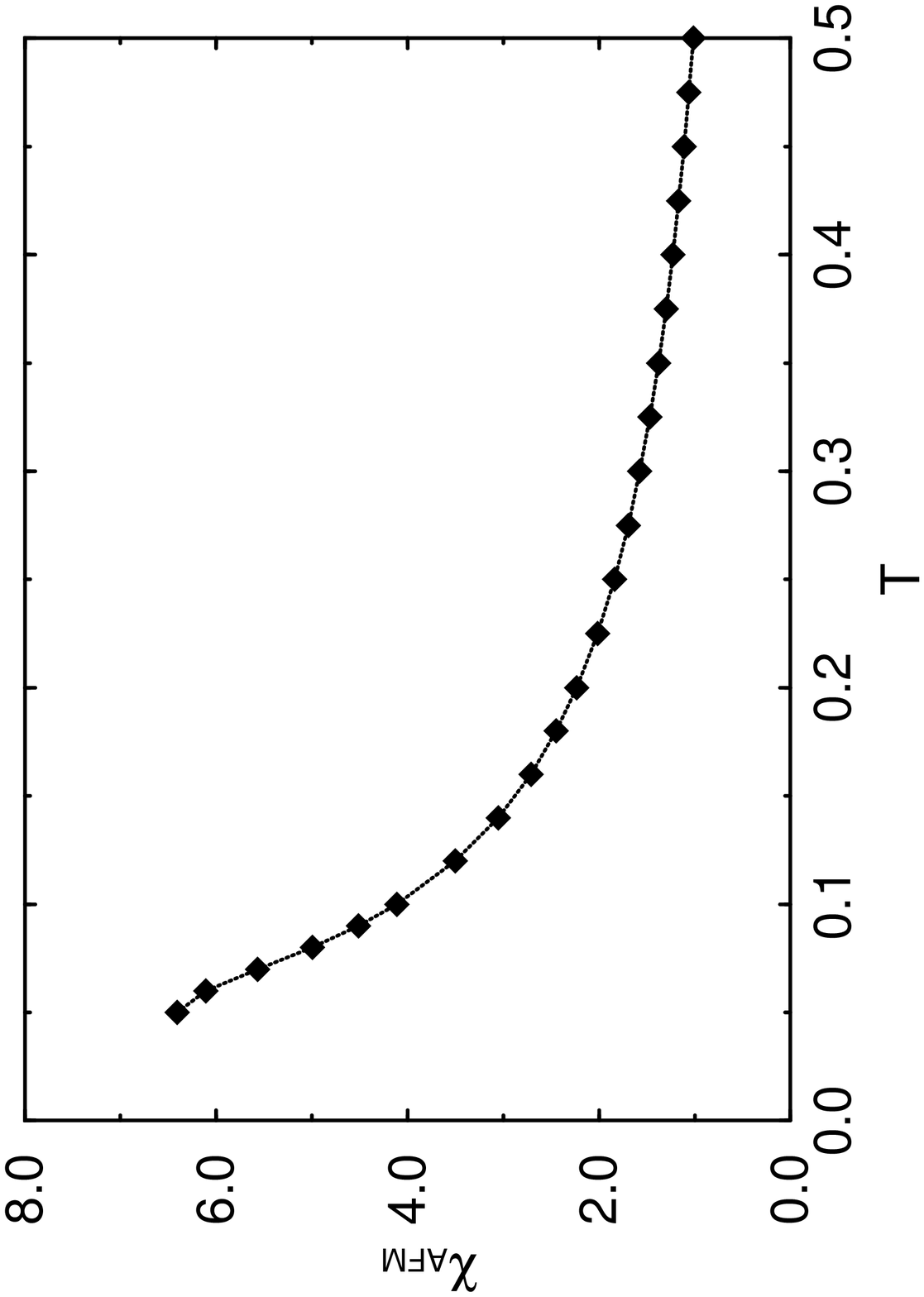}
\center{
\parbox{5.5in}{
{\small
\caption{
    The fully self-consistent spin response for
    ${\bf q}= {\bf Q} = (\pi, \pi)$. Note that
    $\chi_{AFM}(T)$ does not display the rapid
    increase with decreasing $T$ that is observed
    in $T_{sf}({\bf Q}, 0)$.
}
\label{fig5}
}
}}
\end{figure}

     We emphasize that the rapid rise in $T_{sf}(Q, 0)$ with decreasing
  temperature need not be reflected in any physically measurable
  response function and does not necessarily signal a phase transition.
  The fully self-consistent staggered spin susceptibility as a
  function of temperature is presented in Fig. 5.
  The staggered spin susceptibility was calculated
  from a trace of the self-consistent Green's functions in the
  presence of an applied staggered field.  The absence in $\chi_{AFM}$
  of any reflection of the rapid rise in  $T_{sf}(Q, 0)$, dramatically
  contrary to expectations from an RPA calculation of the response, shows
  the importance of vertex corrections in the FEA response
  functions near an instability.
\begin{figure}[htb]
\vspace{4.520in}
\includegraphics{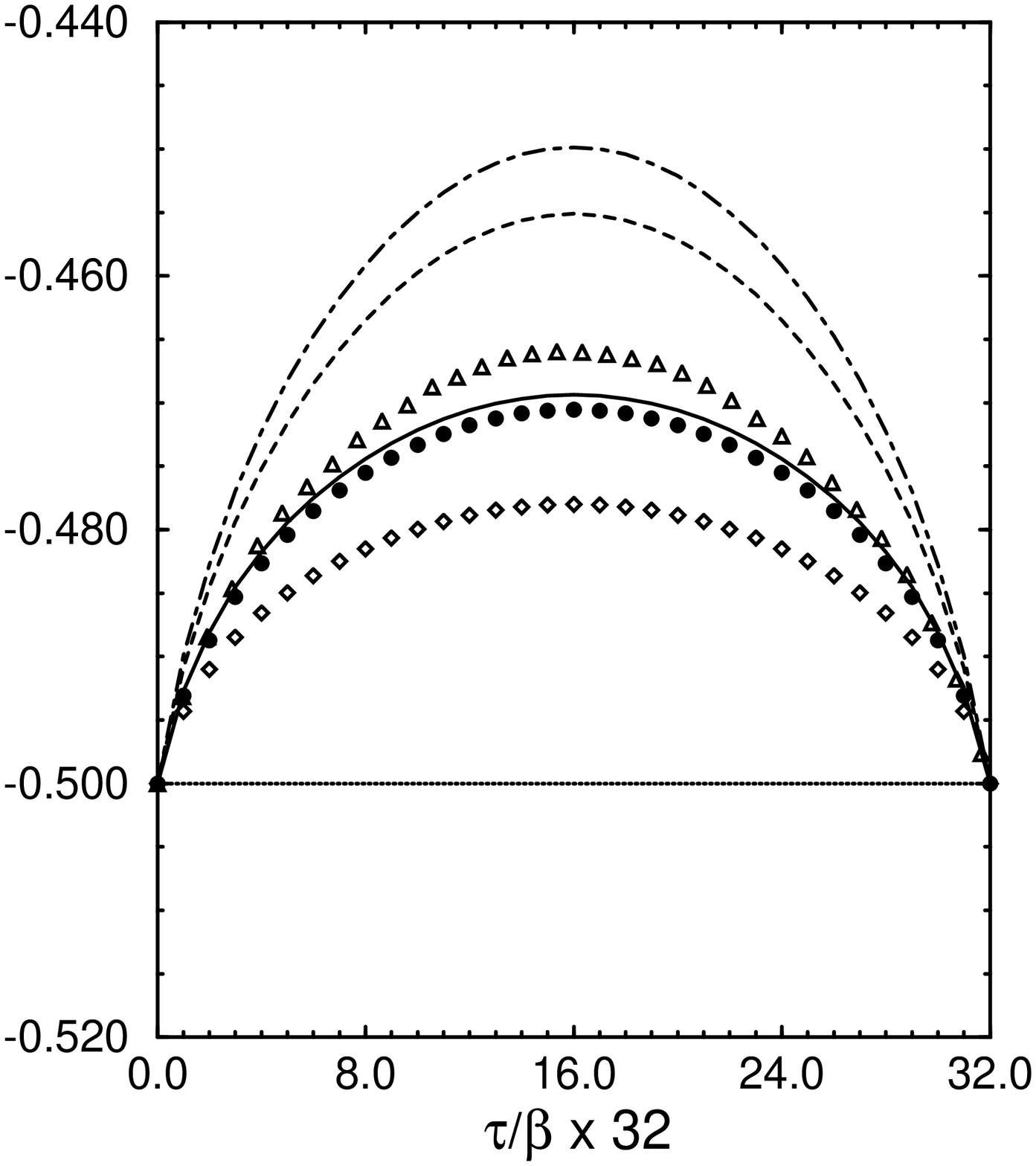}
\center{
\parbox{5.0in}{
{\small
\caption{
   The single-particle Green's function
    $G({\bf k}= \pi/2, \tau)$ for an eight-site
    half-filled Hubbard chain for $U = t$ and $T=0.20$
    calculated in
    the parquet approximation (solid) compared
    with the bare propagator (dotted), quantum
    Monte Carlo ($\triangle$), and the
    single-particle self-consistent perturbation
    theories: the second order diagram alone ($\bullet$),
    the T approximation ($\diamond$),
    the shielded potential approximation (dot-dash),
    and the fluctuation exchange approximation (dashed).
    Note that of the self-consistent perturbation theory
    calculations, the parquet approximation agrees
    most closely with the quantum Monte Carlo result.
}
\label{fig6}
}
}}
\end{figure}

 \begin{center}
  {\large {\it \S 5 Beyond FEA: Calculations in Parquet Approximations}}
 \end{center}

   De Dominicis (De Dominicis, 1963) reformulated the propagator
   renormalized perturbation theory of Luttinger and Ward to
   eliminate not only the single-particle potentials in favor
   of single-particle Green's function but also to eliminate
   the two-body potential in favor of the two-particle propagator
   $G_2$ (or equivalently the two-particle vertex function).
   In this theory, the entropy $S$ plays a role analogous to that of
   the grand potential in the theory of Luttinger and Ward;
   the entropy may be viewed as a functional of
   $G$, $\Sigma$, $G_2$ and the two-particle vertex function $\Gamma$.
   The stationarity conditions for $S$ with respect to variations of $G$,
   $\Sigma$, $G_2$ and $\Gamma$ constitute a closed set of equations
   that must be solved self-consistently. Here we present the equations
   for the two-particle self-consistent perturbation theory
   schematically and refer the interested reader to the original
   works for the full formulation. These are the: Dyson's equation for
   the single particle propagator in Eq. (\ref{dyson}); a
   relation between the self-energy, the bare interaction $U$,
   and the renormalized two-particle vertex
   function
   \begin{equation}
   \Sigma = UGGG \Gamma,
   \end{equation}
   which includes implict double integrals over momenta and double
   sums over frequency; the decomposition
   of the vertex function in terms of a fully two-particle irreducible
   vertex $V$ and $T$-matrices in particle-hole, crossed particle-hole,
   and particle-particle channels,
   \begin{equation}
   \Gamma = V + T_{ph} + \tilde{T}_{ph} + T_{pp};
   \end{equation}
   and a set of Bethe-Salpeter equations relating a $T$-matrix in the $i$-th
   channel to a vertex function $I_i$ that is two-particle irreducible
   in that channel
   \begin{equation}
   T_i = I_i GG I_i + I_i GG T_i.
   \end{equation}
   This set of equations is closed by expressions relating
   each vertex $I_i$ to the $T$-matrices and $V$.
   The resulting theory is self-consistent at the level
   of the single- and two-particle propagators.

   We have solved these equations on the CM-5, using a parallel
   algorithm we describe elsewhere (Hess and Serene, 1995), for
   small Hubbard chains.  We employ two additional approximations
   in our parquet approximation calculations
   that are outside the framework of the exact theory:
   the fully two-particle irreducible vertex $V$ is taken
   to be the bare Hubbard interaction, and  a high
   frequency cutoff is imposed with periodic boundary conditions in
   discrete frequency space (see Deisz, {\em et al.}, 1994).  Both of these
   approximations may be relaxed (Hess and Serene, 1995).

   In Fig. 6, we show the single-particle propagator $G({\bf k}, \tau)$
   calculated in the parquet approximation described above, together
   with $G({\bf k}, \tau)$ from various single-particle self-consistent
   approximations including the FEA, the shielded potential approximation
   (includes only the particle-hole bubble diagrams
   together with the second order diagram),  the T-approximation
   (includes only particle-particle ladder diagrams together with
   the second order diagram), and an approximation
   including only the second-order diagram.  The self-consistent
   calculations were all performed using the same high frequency
   cutoff scheme.  For comparison we show
   $G({\bf k}, \tau)$ obtained from
   quantum Monte Carlo (Assaad, 1995).  Because of the increased
   importance of fluctuations in 1D, we expect that the FEA,
   shielded potential approximation and T-approximations to be poor.
   Of all the diagramatic approaches shown here, the parquet approximation is
   closest to the quantum Monte Carlo result for a $U$ equal to one
   quarter of the bare bandwidth.
   In agreement with our sum-rule arguement in \S 3,
   the FEA provides a better approximation than
   shielded potential approximation. The
   approximation containing only the second-order `skeleton'
   diagram is in better agreement with QMC than any of the other
   single-particle self-consistent theories, at least for this
   relatively small $U$.  This may in part
   reflect the better agreement with the second moment sum-rule
   on the single-particle spectral function.

 \begin{center}
  {\large {\it \S 6 Conclusions}}
 \end{center}

   We have shown that the coupling of spin-fluctuations to
 quasiparticles, as both evolve with decreasing temperature,
 leads to the formation of a non-Fermi liquid state in the
 2D Hubbard model at half-filling in the fluctuation exchange
 approximation. The non-Fermi liquid state shows a positive
 slope in the real part of sigma and a local minimum in the
 imaginary part at zero energy. The energy scale over which
 these anomalies extend is roughly the size of the zero
 temperature gap in mean field theory.
 We have shown that this behavior is preceded
 by the evolution of a sharp peak the spin-fluctuation $T$-matrix
 at ${\bf Q}=(\pi, \pi)$ and that the evolution of this structure
 continues with decreasing temperature. An analysis of the
 $T$-matrix suggests the existence of two temperature regimes
 separated by a characteristic temperature $T^*$. For
 temperatures above $T^*$, the self-energy does not show
 the anomaly; for temperatures below $T^*$, the anomaly in
 the self-energy is present and the sharp structure in
 the $T$-matrix can be described by a single temperature
 dependent energy scale $\omega^*$ which decreases rapidly
 with decreasing temperature.  While the FEA may provide
 a useful model for coupled quasiparticles and spin-fluctuations
 that self-consistently includes the effects
 of evolving low-energy anomalies, it is both desirable and
 possible to go beyond the FEA. We presented calculations that are
 self-consistent at the levels of both the single- and
 two-particle propagators for the Hubbard model in 1D.
 Two approximations were employed: replacing the
 fully two-particle irreducible interaction by the bare interaction,
 and imposing a high frequency cutoff.  Our results for
 the single particle propagator suggest the importance of
 vertex corrections in calculating the self-energy and
 compare favorably with quantum Monte Carlo calculations.
 In future work we will explore the sensitivity of our results
 to the structure of the fully two-particle irreducible
 interaction and discuss the structure of the vertex function.

  \begin{center}
  {\large  Acknowledgements}
 \end{center}

  We thank F. Assaad for supplying quantum Monte Carlo results
  for the 1D Hubbard model.
  It is a pleasure to acknowledge useful discussions with
  A. Kampf,  A. Millis, and D. Pines.
  One of us (DH) would like to thank Lisa Noordergraaf, Bob Weisbeck
  and Mike Young at the NRL CMF for useful discussions.
  This work was supported in part by a grant of
  computer time from the Office of Naval Research and the
  DoD HPC Shared Resource Centers: Naval Research Laboratory
  Connection Machine facility CM-5; Army High Performance
  Computing Research Center under the auspices of Army Research
  Office contract number DAAL03-89-C-0038 with the University
  of Minnesota.

 \begin{center}
  {\large References}
 \end{center}

\begin{list}{ }{\leftmargin 0.0in \itemsep 0.0in}

 \item Anderson, P.W., 1987, {\it Science} {\bf 235}, 1196.

 \item Anderson, P.W., 1990, {\it Phys. Rev. Lett.} {\bf 64}, 1839.

 \item Assaad, F., 1995, private communication.

  \item Baym, G, and Pethick, C.J., 1991,
        {\em Landau Fermi-Liquid Theory},
        (Wiley, New York).

  \item Berk, N., and Schrieffer, J.R., 1966,
        {\it Phys. Rev. Lett.}{\bf 17}, 433.

  \item De Dominicis, C. N., 1963, {\it J. Math. Phys.} {\bf 4}, 255.

  \item Dagotto, E., 1994, {\it Rev. Mod. Phys.} {\bf 66}, 763.

  \item Deisz, J.J., Hess, D.W., and Serene, J.W., 1994,
   {\it Recent Progress in Many-Body Theories}, vol. 4,
   edited by E. Schachinger, {\it et al.}, Plenum, New York.

  \item Deisz, J.J., Hess, D.W., and Serene, J.W., 1995,
  {\em Bull. Am. Phys. Soc.} {\bf 40}, 504;
  Deisz, J.J., Hess, D.W., and Serene, J.W., 1995,
  preprint.

  \item Deisz, J.J., Serene, J.W., and Hess, D.W., 1995,
   in preparation.

  \item Doniach, S. and Engelsberg, S., 1966,
        {\it Phys. Rev. Lett.}{\bf 17}, 750.

 \item Hess, D.W. and Serene, J.W., in preparation.

 \item Hirsch, J.E., and Tang, S., 1989, {\it Phys. Rev. Lett.} {\bf 62}, 591.

 \item Kampf, A.P., 1994, {\it Phys. Rep.} {\bf 249}, 219.

  \item Luttinger, J.M., and Ward, J.C., 1960,
       {\it Phys. Rev.} {\bf 118}, 1417.

  \item Millis, A., Monien, H., and Pines, D., 1990, {\it Phys. Rev.}
    {\bf B} 42, 167.

  \item Pines, D., 1990, {\it Physica} {\bf B 163}, 78.

  \item Serene, J.W., and Hess, D.W., 1992,
   {\it Recent Progress in Many-Body Theories}, vol. 3,
   edited by T.L. Ainsworth {\em et al.}, Plenum, New York.

   \item White, S.R., Scalapino, D.J., Sugar, R.L., Loh, E.Y.,
    Gubernatis, S.E., and Scalettar, R.T., 1989,
    {\it Phys. Rev.} {\bf B 40}, 506.

  \item Vidberg, H.J., Serene, J.W., 1977, {\em J. Low Temp. Phys.},
  {\bf 19}, 179.

\end{list}

 \end{document}